\documentclass[prb,aps,twocolumn,superscriptaddress,floatfix,citeautoscript,eqsecnum,groupedaddress,nobalancelastpage,showpacs]{revtex4-1}
\usepackage{graphicx,amsmath,amsfonts,amssymb}
\usepackage[T1]{fontenc}
\usepackage{hyperref}
\usepackage{color}

\begin{document}
\bibliographystyle{apsrev}

\title{Hydrodynamic spin fluctuations in the antiferromagnetic Heisenberg chain}

\author{Yousef Rahnavard} 
\email{y.rahnavard@tu-bs.de} \affiliation{Institute for Theoretical
Physics, Technical University Braunschweig, D-38106 Braunschweig,
Germany}

\author{Robin Steinigeweg} 
\email{r.steinigeweg@tu-bs.de} \affiliation{Institute for Theoretical
Physics, Technical University Braunschweig, D-38106 Braunschweig,
Germany}

\author{Wolfram Brenig} 
\email{w.brenig@tu-bs.de} \affiliation{Institute for Theoretical
Physics, Technical University Braunschweig, D-38106 Braunschweig,
Germany}

\date{\today}

\begin{abstract}

We study the finite temperature, low energy, long wave-length spectrum of
the dynamic structure factor of the spin-$1/2$ antiferromagnetic Heisenberg
chain in the presence of exchange anisotropy and external magnetic fields. Using
imaginary-time quantum Monte-Carlo we extract parameters, relevant to
characterize a {\it renormalized} Luttinger liquid. For small momentum our
results are consistent with a change from propagating spinon density waves
to spin diffusion, described by a finite-frequency spin-current relaxation
rate. Results for this relaxation rate as well as other Luttinger liquid
parameters are presented versus temperature, momentum, magnetic field,
and anisotropy, including finite-size analysis, and checks for anomalous
diffusion. Our results are consistent with exact diagonalization and Bethe
Ansatz, where available, and with corroborate findings of other previous studies
using bosonization, transfer matrix renormalization group, and quantum
Monte-Carlo.
\end{abstract}
\maketitle

\section{Introduction}

The one-dimensional (1D) spin-$1/2$ antiferromagnetic Heisenberg chain
(AFHC) is one of the most fundamental models of quantum many-body physics.
It is relevant to low-dimensional magnets \cite{johnston2000}, ultra-cold
atoms \cite{trotzky2008}, nanostructures \cite{gambardella2006}, and --
seemingly unrelated -- fields such as string theory \cite{kruczenski2004}
and quantum Hall systems \cite{kim1996}. In the presence of an external
magnetic field, its generalization to anisotropic exchange, the XXZ model,
reads
\begin{equation}
H=J \sum_{l=1}^L [\frac{1}{2}(S_{l}^{+}S_{l+1}^{-}{+}S_{l}^{-}S_{l+1}^{+}){+}\Delta
S_{l}^{z}S_{l+1}^{z}{-}hS_{l}^{z}]\,,\label{1}
\end{equation}
where $J>0$ is the antiferromagnetic exchange interaction with an anisotropy
ratio $\Delta$, $S^{\pm,z}_l$ are the spin operators on site $l$ of a
chain of length $L$ with periodic boundary conditions (PBC), and $B=g\mu_B\hbar h$ is the magnetic field.

Experimentally, dynamical correlation functions of the AFHC have recently
become accessible to a variety of high-resolution spectroscopies at finite
temperature and in the presence of external magnetic fields, e.g.\ inelastic
neutron scattering (INS) \cite{Stone2003a,Lake2005a,Mourigal2013a},
high-field nuclear magnetic resonance (NMR)
\cite{Takigawa1997a,Thurber2001a,Wolter2005a,Kuhne2008a}, muon
spin-resonance ($\mu$SR) \cite{Pratt2006a}, and magnetic transport
\cite{Hess2007a,HeidrichMeisner2007a,Sologubenko2007a}.

Theoretically, and while the AFHC is integrable, its dynamical spin
correlation functions remain a major challenge. Analytically, significant
progress has been made at low temperatures by calculating multi-spinon
response functions \cite{Caux2005-12} from Bethe Ansatz, including the
cases of $h\neq 0$ and $\Delta\neq 1$. Important insight has also been
obtained in the continuum limit and at low temperatures by bosonization
\cite{Pereira2007a,Pereira2012a}. Perturbation theory allows to access
regimes of $\Delta\gg 1$ \cite{James2009a}. Numerically, at finite
temperature, time-dependent density-matrix renormalization group (t-TDMRG)
\cite{Barthel2009a} is a very powerful approach. However, at present,
entanglement growth remains a limiting factor on the time window to access
low-energy, long-wavelength dynamics
\cite{Karrasch2012a,Karrasch2013a}. Recently, dynamical quantum typicality
(t-QT) has been shown to overcome such limitations
\cite{Elsayed2013a,Steinigeweg2014a,Steinigeweg2014b}, however, only at high
temperatures. Early on, exact diagonalization
\cite{Fabricius1997a,Werner2001a,Waldmann2001a} and, more recently, Lanczos
variants \cite{Herbrych2012a} have been used. However, finite-size or
Krylov-space-dimension effects limit their spectral resolution.

Quantum Monte-Carlo (QMC) is an additional complementary approach for the
HAFC.  It is applicable from low to high temperatures and allows to
consider systems almost in the thermodynamic limit, including finite
magnetic fields and anisotropy. {\it Static} correlation functions can be
obtained from it with arbitrary precision, however, the evaluation of spin
{\it spectra} from QMC
\cite{Deisz1990a,Deisz1993a,Starykh1997a,Grossjohann2009a} requires
analytic continuation of imaginary-time data, leading to errors from
maximum-entropy approaches \cite{Jarrell1996}. This renders the evaluation of
the dynamic structure factor $S(q,\omega)$ by QMC very challenging, in
particular for small momenta $q\ll 1$, where sharp low-energy spectral
features are expected. Recently therefore, QMC results
\cite{Grossjohann2010a} for $S(q,\tau)$ at imaginary time $\tau$ have been
compared directly to suggestions from bosonization and time-dependent
transfer-matrix renormalization group (t-TMRG)
\cite{Sirker2009a,Sirker2011a}, corroborating a picture of finite-temperature
{\it diffusive} spin dynamics at finite frequencies in the
low-energy, long-wavelength limit. Such spin diffusion in the HAFC is a
long-standing issue \cite{Steinigeweg2011a}, relating the magnetization dynamics
to the question of dissipation of spin currents and the quest for a
spin-Drude weight in the XXZ model, see e.g.\ Ref.\ \onlinecite{Steinigeweg2014a}
and references therein.

In this context, the aim of the present work is to extend the information
obtained from the QMC approach of Ref.\ \onlinecite{Grossjohann2010a} into various
directions, including physical as well as technical aspects. The manuscript
is organized as follows. In Sec.\ \ref{method}, we describe our method of
analysis.  In Sec.\ \ref{isochn} we revisit Ref.\ \onlinecite{Grossjohann2010a}
from three directions, first, by subjecting its findings to a finite-size
scaling analysis, second, by comparing our extended results with exact
diagonalization, and third, by checking for anomalous corrections to
diffusion. The following Secs.\ \ref{flddep} and \ref{Aispy} extend the
analysis in two additional ways, namely, by studying the impact of finite
magnetic fields and anisotropy. Section \ref{chainsum} summarizes our findings.

\section{QMC Approach}\label{method}

The prime object of interest in this work is the Fourier transform of the
{\it longitudinal} dynamic structure factor

\begin{equation}
S(q,\omega) =
\int_{-\infty}^{\infty} dt \, e^{i\omega t} S(q,t)
\end{equation}
at small momentum $q$ and frequency $\omega$, and its corresponding retarded
dynamical spin susceptibility $\chi_{\text{ret}}(q,\omega) \equiv \chi^\prime+i
\chi^{\prime\prime}$ is related to $S(q,\omega)$ by the
fluctuation-dissipation theorem $\chi^{\prime\prime}(q,\omega) =
[1-\exp(-\beta\omega)]S(q,\omega)/2$ at inverse temperature $\beta = 1/T$. Here $t$ refers to real time and
$S(q,t)=\langle S^z_q(t)S^z_{-q}\rangle$ with $S^z_q=\sum_{l} e^{-i q l}
S^z_l$ being the spin z component at momentum $q$. 

$S(q,\omega)$ is related to the imaginary-time structure factor $S(q,\tau)
= \langle S^z_q(\tau)S^z_{-q}\rangle$ by analytic continuation through the
usual integral transform $S(q,\tau)=\int_{0}^{\infty}d\omega K(\omega,
\tau)S(q, \omega)/\pi$ with a kernel $K(\omega, \tau)=e^{-\tau\omega} +
e^{-(\beta-\tau)\omega}$. While QMC allows to evaluate $S(q,\tau)$ with
high precision, inverting the latter integral transform, e.g.\ by maximum
entropy methods \cite{Jarrell1996}, is mathematically ill-posed and
introduces errors which can be considerable in particular at small $q$ and
$\omega$ \cite{Grossjohann2009a}.  A central point of the present work is,
that instead of performing such analytic continuation, we follow
\cite{Grossjohann2010a} and express $\chi(q,\omega)$ through only a few
relevant parameters, by fitting numerical imaginary-time QMC data to an
educated analytic guess for $\chi(q,\tau)$. While this avoids errors from
analytic continuation, it obviously represents a bias which requires {\it
posteriori} quality checks.

Throughout this paper we will be interested in the long-wavelength, low-energy
limit, and in a temperature range where we may assume that the physics of
Eq.\ (\ref{1}) can be described by a {\it renormalized} Luttinger liquid
(LLQ). Therefore we start from
\begin{equation}
 \chi_{\text{ret}}\left(q,\omega\right)=
\frac{Kvq^{2}/2\pi}{\omega^{2}-v^{2}q^{2}-\Pi_{\text{ret}}(q,\omega)}\,.
\label{Pi2ndorder} 
\end{equation}
Except for $\Pi_{\text{ret}}$, this form of $\chi_{\text{ret}}$ is dictated by the
response of a {\it free} LLQ with Luttinger parameter $K$ and
spinon velocity $v$. All deviations from the free LLQ are encapsulated in the
irreducible density self-energy $\Pi_{\text{ret}}(q,\omega)$. Causality requires
that the real(imaginary) part of the latter is an even(odd) function in
$\omega$. Assuming that $\Pi_{\text{ret}}(q,\omega)$ is non-singular for small
$\omega$, and since we are interested in the low-$\omega$ limit, we
expand to $O(\omega^2)$
\begin{equation}
\Pi_{\text{ret}}(q,\omega) \simeq u_0(q) + i u_1(q)\omega + u_2(q)\omega^2 +\ldots
\label{PiW1} 
\end{equation}
with $u_i(q)\in {\mathbb R}$. Expanding also to $O(q^2)$ for $q\ll 1$,
inversion symmetry forces
\begin{equation}
\Pi_{\text{ret}}(q,\omega) \simeq c v^2 q^2-2i\gamma\omega-b\omega^2+\ldots
\label{PiW2} 
\end{equation}
with {\it real} constants $c$, $\gamma$, and $b$, and all terms of order
higher than $\omega^2$, $q\omega$, and $q^2$ have been dropped.  Since we
will confine ourselves to $|\Delta|\leq 1$ and the longitudinal dynamical
structure below the saturation field $h\leq h_c = 2$ we may discard terms
constant in $q$ and $\omega$ in Eq.\ (\ref{PiW2}), since they would lead to
a gap in $\chi_{\text{ret}}(0,\omega)$.

We emphasize, that Eq.\ (\ref{PiW2}) follows solely from causality,
analyticity, and symmetry. Therefore, bosonizing the HAFC and performing
proper perturbation theory beyond the free LLQ {\it must} result in exactly
this small $q, \omega$ expansion for $\Pi_{\text{ret}}(q,\omega)$. This is
consistent with
Refs.\ \onlinecite{Pereira2007a,Sirker2009a,Sirker2011a,Pereira2012a}, where
explicit expressions for $c$, $\gamma$, and $b$ have been obtained to
second order in Umklapp scattering and first order in band curvature. 
In contrast to this perturbation theory however, the spirit of our work is very
different. In fact we will view $v$, $K$, $c$, $\gamma$, and $b$ as
{\it parameters} to be determined by fitting to imaginary-time QMC results for
$\chi(q,\tau)$. This is a non-perturbative approach. To this end we
map Eqs.\ (\ref{Pi2ndorder}), (\ref{PiW2}) onto bosonic Matsubara frequencies
$\omega_{n}=2\pi n T$ with integer $n$,
\begin{equation}
 \chi(q,\omega_{n}) =  \frac{K_{q}v_{q}q^{2}/(2\pi)}{(1+b)
\omega_{n}^{2}+(1+c)v_{q}^{2}q^{2}+2\gamma_{q}|\omega_{n}|} \,\, ,
\label{mats}
\end{equation}
and transform this to imaginary time with
\begin{equation}
 \chi\left(q,\tau\right) = 2\sum_{n=0}^{\infty}
\cos(\omega_{n}\tau)\chi(q,\omega_{n})-\chi(q,0) \,\, ,
\label{eq:chiimag}
\end{equation}
Several comments are in order. First, the static susceptibility
$\chi_q=\chi_q(q,0)$ of the XXZ model is known to vary with $q$, e.g., at
$\Delta=1$ and $h=0$ it increases monotonously as
$q\rightarrow\pi/2$. However, $\chi_q=K/[2\pi v(1+c)]$ resulting from
Eqs.\ (\ref{Pi2ndorder}), (\ref{PiW2}) is momentum-independent. To fix this
shortcoming of the free LLQ theory, we allow for an additional momentum
dependence $K\rightarrow K_{q}$, $v\rightarrow v_{q}$ in Eq.\ (\ref{mats}) -
albeit weak at $q\ll1$ -, when matching Eqs.\ (\ref{Pi2ndorder}),
(\ref{PiW2}) with QMC. Second, the constants $b$ and $c$ in Eq.\ (\ref{PiW2})
are redundant when fitting to QMC because they can always be absorbed into
a renormalization of $K_q$, $v_q$, and $\gamma$. Without loss of generality
we therefore fix $b$ and $c$ \cite{fixbc} to the values obtained from the
perturbation theory of
Refs.\ \onlinecite{Pereira2007a,Sirker2009a,Sirker2011a,Pereira2012a}. Third, as
an additional generalization of Eq.\ (\ref{PiW2}) we allow for a momentum
dependence $\gamma \rightarrow \gamma_q$. Fourth, Eq.\ (\ref{Pi2ndorder})
with Eq.\ (\ref{PiW2}) does not capture the finite width of the spectral
function $\chi''(q,\omega)$ at $T=0$, resulting primarily from the
two-spinon continuum. However, at $q/\pi\ll1$ the latter width is of order
$Jq^{3}$, which is negligible against $\gamma_q$ for those wave vectors and
temperatures which we will be interested in.

In summary, our procedure consists in evaluating $\chi(q,\tau)$ by QMC and
fitting the result at each $q$, using Eq.\ (\ref{eq:chiimag}), in terms of
three numbers: $K_q$, $v_q$, and $\gamma_q$.

\subsection{Spin Diffusion}\label{diffsec}

For finite $\gamma_q$ and $\omega\ll\gamma_q$, Eq.\ (\ref{Pi2ndorder})
displays a diffusion pole with a diffusion constant $\Gamma_q = (1+c)
v_q^{2} /(2 \gamma_q)$. Therefore the prime effect of the renormalization
$\Pi_{\text{ret}}(q,\omega)$, which we study in this work, is to generate a
hydrodynamic regime in which the magnetization dynamics of the HAFC is
diffusive. This also has consequences for the {\it regular} spin
conductivity $\sigma_{reg}(\omega)$, which we sketch
briefly next.

\begin{figure}[tb]
\includegraphics[width=0.95\columnwidth]{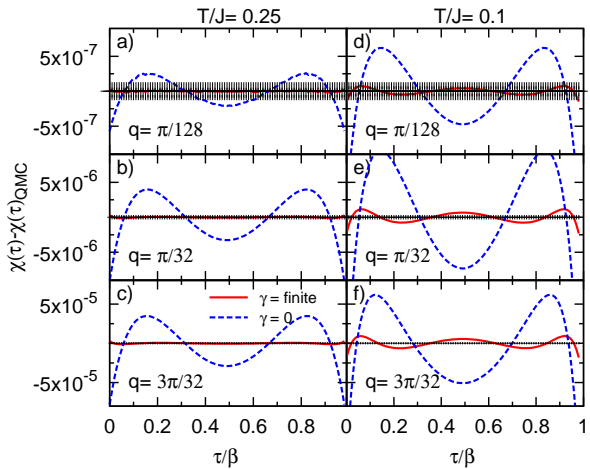}	
\caption[Fitting error at the isotropic point]
{(color online) Differences between the fitted curves and the original QMC results at the
isotropic point for different wave vectors and temperatures.  The
system-size considered here is $L=256$. Error bars shown in the figure indicate the statistical noise of 
the QMC data.}
\label{fig:diffD1h0N256}
\end{figure}

The spin current is defined by the lattice version of the continuity equation
$\partial_{t} S_{q}^{z} = [1-\exp(-iq)] \, j_{q}$, and its conductivity
$\sigma(\omega) =(1{-}e^{\beta\omega}) \mathrm{Re} \int^{\infty}_{-\infty}
e^{i\omega t}\langle j_0(t) j_0 \rangle dt/\omega$ can be decomposed as
\begin{equation}
\sigma(\omega)=D\delta(\omega)+\sigma_{\text{reg}}(\omega)\,,
\label{2} 
\end{equation}
where $D$ is the enigmatic spin Drude weight and $\sigma_{\text{reg}}(\omega)$ is
the regular part of the spin conductivity. The Drude weight of the HAFC has
been of interest for more than two decades by now. We refer to
Ref.\ \onlinecite{Steinigeweg2014a} for the recent status. The regular
conductivity is related to the spectrum $\chi^{\prime\prime}(q,\omega)$
trough the continuity equation
\begin{equation}
\sigma_{\text{reg}}(\omega)=\lim_{q\rightarrow0}\frac{\omega}{q^{2}}
\chi''(q,\omega)\,.
\label{7}
\end{equation}
If $\chi(q,\omega)$ exhibits a diffusion pole in the frequency domain, then in the time domain Eq.\ \ref{7} 
implies exponential decay of the
current on a time scale $1/\gamma_0$ given by the current relaxation rate $\gamma_0$ \cite{exception}.
In the frequency domain it implies a Lorentzian line shape of
$\sigma_{\text{reg}}(\omega)$ with width $\gamma_0$. We emphasize that
Eq.\ (\ref{7}) provides {\it no} insight into the value of $D$. Coexistence of
a diffusive regular conductivity with a finite Drude weight has been
suggested based on bosonization, t-TMRG, and memory-function methods
\cite{Sirker2009a,Sirker2011a}. But also strong momentum dependence beyond
Eq.\ (\ref{PiW2}) has been invoked to question this, based however on small
systems at high temperatures \cite{Herbrych2012a}. For the remainder of
this work we will refrain from speculations on $D$.

\section{Isotropic chain}\label{isochn}

In this section we extend previous work \cite{Grossjohann2010a}, based on
the approach described in Sec.\ \ref{method}, into several new directions.
First we study the effects of system-size. Then we perform a consistency
check of our approach by comparing to available results from exact
diagonalization. Finally, we test if the current relaxation rate shows
relevant energy dependence.

\subsection{System-size and momentum dependence}

In Ref.\ \onlinecite{Grossjohann2010a}, the approach described in the previous
section has been applied to the isotropic HAFC, i.e.\ at $\Delta=1$ and
$h=0$. That work has proven feasibility of the method. In particular it was
shown, that $K_0$ and $v_0$ extracted from it agree very well with
thermodynamic Bethe Ansatz \cite{Lukyanov98,Johnston2000a}, as well as that
the relaxation rate $\gamma_0$ obtained was consistent with that from
bosonization and t-TMRG \cite{Sirker2009a,Sirker2011a} to within factors of
2. We will not repeat this analysis here. Instead we note that the findings
of Ref.\ \onlinecite{Grossjohann2010a} where based on a {\it single} system-size,
i.e.\ $L=128$, and it remained unclear if parts of its results, e.g.\ the
momentum dependencies observed for $\gamma_q$ were finite-size
effects. Therefore, in this subsection, we will analyze $\gamma_q$ versus
system-size.

\begin{figure}[!t]
\includegraphics[width=0.95\columnwidth]{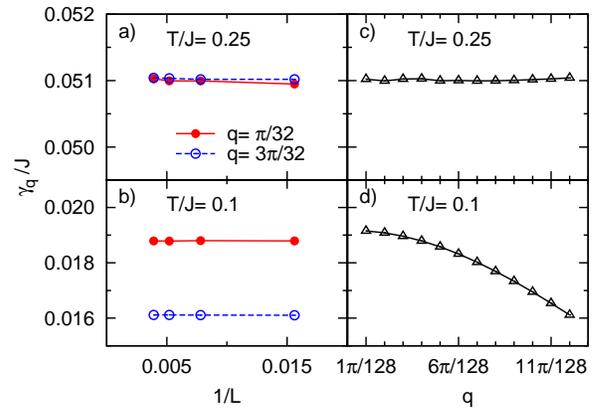}
\caption[Momentum and size dependence of $\gamma_q$]
{(color online) Finite-frequency spin-current relaxation rate $\gamma_q$ of the isotropic
Heisenberg spin chain versus inverse system-size $1/L$ and momentum $q$
for chain lengths $64<L<256$ at two temperatures $T/J=0.1$, $0.25$. For
momentum dependence of $\gamma_q$, the largest system-size $L=256$ is
chosen. 
}
\label{fig:gamma-momentum} 
\end{figure}

To begin we comment on the accuracy of the fitting procedure summarized in
the last paragraph before Sec.\ \ref{diffsec}. To this end
Fig.\ \ref{fig:diffD1h0N256} shows the minimum difference between our QMC
results and Eq.\ (\ref{eq:chiimag}) for optimized parameters $K_q$, $v_q$, and
$\gamma_q$, for different temperatures and small, nonzero wave vectors on a
system with 256 sites. In addition each panel also shows this difference
assuming that $\gamma_q=0$, i.e.\ discarding spin diffusion. The figure also
displays the statistical noise of the QMC data as error bars. Several points
have to be mentioned. First, the quality of our imaginary-time QMC results
is worth noting by observing the absolute order of magnitude on the $y$ axes
of this figure. Second, for all panels shown, including a finite $\gamma_q$
leads to clearly better agreement with QMC than setting $\gamma_q=0$. We
emphasize, that $\gamma$ is $\sim O(J/20\ldots 100)$, which implies that our
method is capable to sense spectral structures on such small energy
scales. This is unlikely for standard maximum-entropy analytic continuation. Third, at
intermediate temperatures, $T=0.25J$, using Eqs.\ (\ref{Pi2ndorder}),
(\ref{PiW2}) with finite $\gamma_q$ is well within the standard deviation of
QMC for all momenta depicted. For lower temperatures this is so at least
for the smallest momenta, i.e.\ as $q\rightarrow 0$. Therefore these plots
also show that, while a finite $\gamma_q$ clearly improves agreement with
QMC, the range of validity in momentum space of the purely hydrodynamic,
diffusive description shrinks as the temperature is lowered. Similar
observations have been made in Ref.\ \onlinecite{Steinigeweg2011a}. Future work
should elaborate on this. Finally, the figure shows that $\gamma_q$ decreases
with temperature \cite{Sirker2009a,Grossjohann2010a,Sirker2011a}.

\begin{figure}[tb]
\includegraphics[width=0.95\columnwidth]{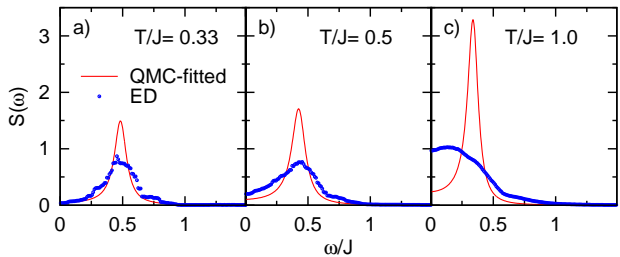}
\caption[ED and QMC spectra]
{(color online) Hydrodynamic spectra in the frequency domain with parameters fitted to QMC
as compared to ED, both for length $L=18$, momentum $q=\pi/9$, the isotropic point
$\Delta = 1$, and three temperatures $T/J=\{0.33,0.5,1.0\}$. 
}
\label{fig:ED}
\end{figure}

Next we turn to the system-size dependence of $\gamma_q$. To that end we
consider four chains with sizes of $L=64,128,192,256$. The results
are shown in Fig.\ \ref{fig:gamma-momentum} a), b) for two different
temperatures $T/J=0.1$, $0.25$ and two momenta. Note that these momenta,
$q=\pi/32$, $3\pi/32$, have been chosen to be identical despite the
differing system-sizes, which implies that they are not necessarily the
smallest wave vector possible for a given $L$.  The main point of this
figure is, that we observe {\it no} finite-size dependence. This is even
more so in view of the $y$ axes scales depicted. Therefore our results can
safely be considered as being in the thermodynamic limit.

In panels c), d) of this figure we show $\gamma_q$ versus momentum for
the largest system, i.e., $L=256$. Obviously, for the lower temperature
$T=0.1$, $\gamma_q$ slightly decreases with $q$ \cite{hTlq}. This is fully
identical to the findings of Ref.\ \onlinecite{Grossjohann2010a}, where however
$L=128$ was used. I.e.\ this decrease is not a finite-size effect as was
speculated in the latter work. Rather, and in view of the deviation in
Fig.\ \ref{fig:diffD1h0N256} e), f), we believe that the crossover from
a constant to a weakly momentum-dependent $\gamma_q$ is another indication
of a shrinking hydrodynamic regime as the temperature is lowered.

\begin{figure}[tb]
\includegraphics[width=0.95\columnwidth]{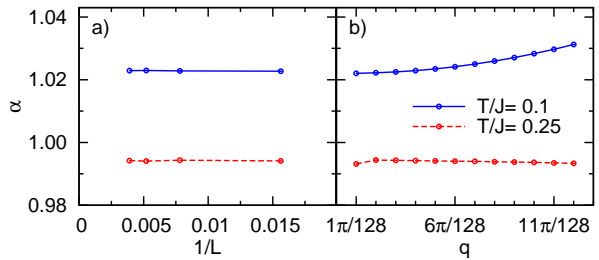}
\caption[Power law frequency dependence of the diffusion kernel]
{(color online) Exponent $\alpha$ of the spin-density relaxation rate $\gamma
\omega^\alpha$ of the isotropic Heisenberg chain versus a) inverse system-size $1/L$ at momentum $q=\pi/32$ of system-sizes $64<L<256$, b)
momentum $q$ for a system-size of $L=256$. In both panels, two temperatures $T/J=0.25$, $0.1$ are considered.}
\label{fig:alpha-momentum}
\end{figure}

\subsection{Comparison with exact diagonalization}

Apart from Fig.\ \ref{fig:diffD1h0N256}, another quality check of our method
can be constructed by comparison with results from exact diagonalization
(ED). While the latter are beyond any doubt, they can only be obtained on
small systems, in particular if complete diagonalization is performed using
a canonical ensemble. Small system-sizes imply that the smallest, nonzero
wave vectors remains rather large, and moreover, that acceptably smooth
spectra can only be obtained for rather large temperatures. Nevertheless, it
is very tempting to compare our QMC approach with ED results.

Figure \ref{fig:ED} compares the evolution with temperature of the
longitudinal structure factor $S(q,\omega)$, obtained from both, the QMC
method of Sec.\ \ref{method} and ED on chains of identical length,
i.e.\ $L=18$, at the smallest, nonzero wave vector $q=\pi/9$. This
comparison is remarkable. First, it is obvious, that as $T$ is lowered, the
agreement between ED and the QMC approach improves monotonously. At the
lowest temperature the width of the two spectra are almost identical. The
peak heights are still different. We attribute this to two effects. 
First, at low temperature respective finite-size effects are more prominent.
Second, the width of the ED spectrum
is still slightly larger than that of QMC. In turn, sum-rule effects will
decrease the ED's peak intensity. As we increase $T$ the line shape of the
ED clearly broadens beyond that of the QMC at $T/J=0.5$, and for $T/J=1.0$ the
agreement is lost.

Exactly this variation with temperature is expected. For low $T$ a
description of the HAFC in terms of a renormalized LLQ should hold,
explaining the rather similar spectra from ED and our QMC approach. We
speculate that this agreement should even improve for $T<0.33J$, where
however ED spectra are too noisy. Increasing the temperature, it is known
that up to $T/J\lesssim 0.1$ perturbatively renormalized LLQs provide a
good description of {\it thermodynamic} properties of the HAFC which agree
rather well with static QMC or TBA \cite{johnston2000,Lukyanov98}, while
for $T/J\gtrsim0.25$ the agreement starts to deteriorate. In view of this
breakdown of the LLQ description, it is natural that ED and our QMC
approach start to differ for $T/J>0.5$, as in Fig.\ \ref{fig:ED}.

\begin{figure}[tb]
\includegraphics[width=0.95\columnwidth]{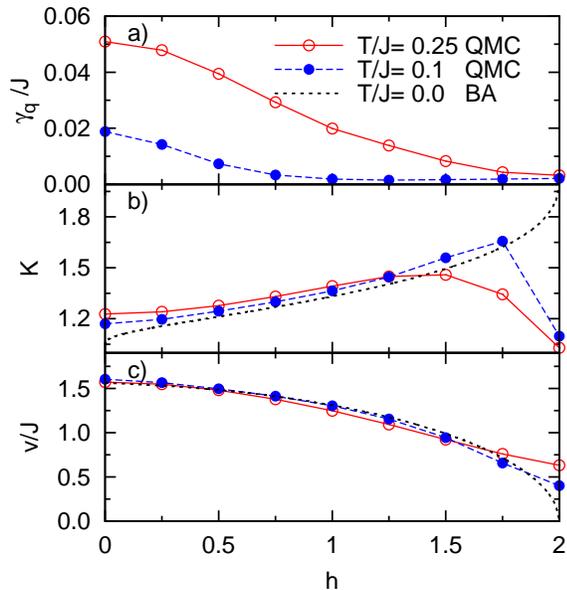}
\caption[Field dependence of $\gamma_q$, $K$, $v$]
{(color online) a) Spin current relaxation rate $\gamma_q$, b) Luttinger parameter $K$, and c) spin
velocity $v$ at finite temperatures $T/J=0.1$, $0.25$  obtained from QMC
of a chain with $64$ sites. Luttinger parameter $K$ and spin velocity
$v$ at zero temperature known from BA in terms of magnetic field are also
shown. In the case of QMC, the momentum is $q=\pi/32$. 
}
\label{fig:gamma-h}
\end{figure}

\subsection{Anomalous diffusion}

As a final check we test the stability of our approach against
modifications of the basic diffusion law by super(sub)diffusion. The latter
is defined by a density which spreads in time according to $\langle
r^2\rangle \sim t^\alpha$ with $\alpha >1(<1)$. In terms of
Eq.\ (\ref{PiW2}), this implies a substitution
\begin{equation}
\gamma\omega \rightarrow \gamma \omega^\alpha
\label{andiff}
\end{equation}
with $\alpha\neq 1$. Recently, anomalous diffusion has been claimed to
result from Lindblad quantum master equations at the isotropic point and
infinite temperatures \cite{Znidaric2011,Znidaric2014a}. This claim
is consistent with numerical simulations of spin-density dynamics on the
basis of classical mechanics \cite{CM}.

Allowing for the modification of Eq.\ (\ref{andiff}) within our approach is
straightforward and leads to one additional fitting parameter. The results
for $\alpha$ at two temperatures $T=0.1J$, $0.25J$ are summarized in Fig.\ \ref{fig:alpha-momentum} .
Panel a) of this figure depicts finite-size scaling of $\alpha$ for system-sizes $64<L<256$ and one momentum $q=\pi/32$.
$\alpha$ versus momentum for a system-size of $L=256$ is shown in panel b) of this figure.
To summarize, $\alpha$ remains very close to one, i.e.\ any
tendencies towards super(sub)diffusion at low temperatures are negligible
and can safely be discarded.  We note in passing, that all modifications we
observe in $K_q$, $v_q$ and $\gamma_q$, due to allowing for $\alpha\neq 1$,
are also negligible.

\begin{figure}[tb]
\includegraphics[width=0.95\columnwidth]{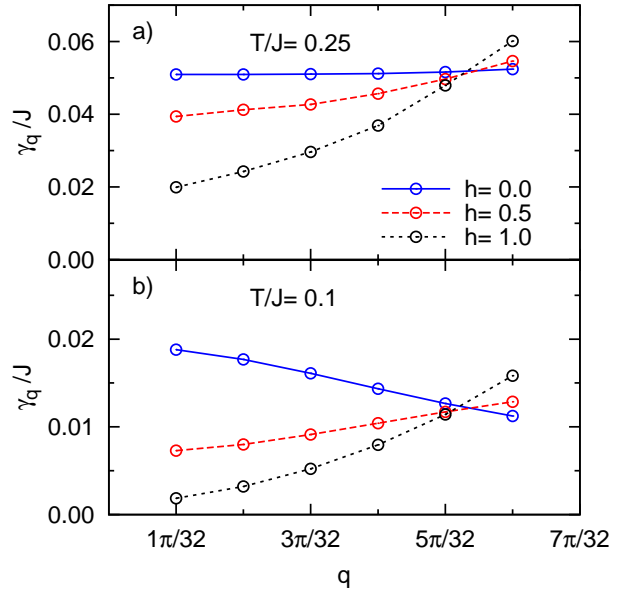}
\caption[Momentum dependence of $\gamma_q$ at finite fields]
{(color online) Momentum dependence of the spin current relaxation rate for two
temperatures a)$T/J=0.25$, b)$T/J=0.1$, and different external magnetic fields is
plotted. The system-size considered here is $L= 64$.
}
\label{fig:gamma-q-h} 
\end{figure}

\section{Magnetic field dependence}\label{flddep}

In this section, we will extend our approach to the case of finite magnetic
fields in the isotropic chain. Since we consider the longitudinal
dynamical structure factor, the spectrum remains gapless below the
saturation field $h_c = 2$.

First, Fig.\ \ref{fig:gamma-h}a) shows $\gamma_q$ versus field for two
temperatures $T=0.1$, $0.25$ $J$ at wave vector $q= \pi/32$ in a system
with $L=64$. This evidences a monotonous decrease of the relaxation rate
with magnetic field. Such a behavior is consistent with predictions at low
fields from bosonization \cite{Sirker2011a}. While overlap between the spin
current and the conserved heat current of the HAFC at nonzero $h$
irrevocably imply a finite spin Drude weight for $0{<}h{<} h_c$
\cite{Mazur1971a,Zotos1997a,FHM2005a}, this allows for no conclusion on the
regular conductivity $\sigma_{\text{reg}}(\omega)$ from Eqs.\ (\ref{2}), (\ref{7}). In
Refs.\ \onlinecite{Sirker2009a,Sirker2011a} this has been dubbed coexistence of
ballistic and diffusive transport channels, meaning that in the time
domain, and on relatively short scales, currents relax diffusively onto a
constant infinite-time limiting value. Therefore, while our nonzero
$\gamma_q$ does not contradict a finite Drude weight, it is nevertheless
reassuring that the decrease of $\gamma_{q_\text{min}}$ we observe is indicative
of a `more' ballistic behavior of the system as $h$ increases.

\begin{figure}[tb]
\includegraphics[width=0.95\columnwidth]{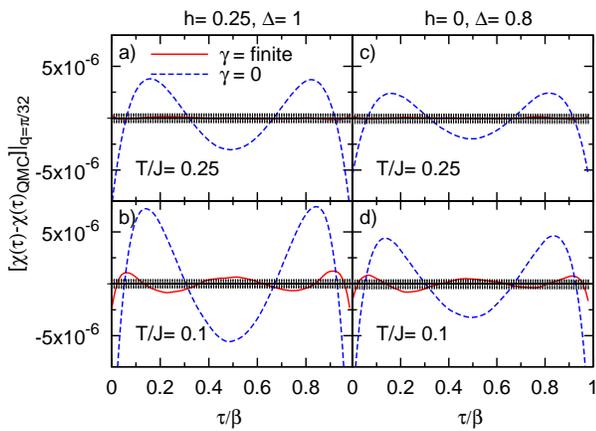}	
\caption[Fitting error at finite fields]
{(color online) Differences between the fitted curves and original QMC
data at momentum $q=\pi/32$ and two temperatures  $T/J=0.25$, $0.1$. In panels a) and b),
magnetic field is set to $h=0.25$ and anisotropy is set to $\Delta=1$ while in panels c) and d), $h=0$ and $\Delta=0.8$.
The system-size for all cases is $L= 64$ and the error bars indicate the statistical noise of the QMC data.
}
\label{fig:diff-h-delta} 
\end{figure}

Next we turn to the remaining two fitting parameters $K_q$ and $v_q$. As
the magnetic field sets the chemical potential of the Jordan-Wigner
fermions, it is clear that apart from changes in $\gamma_q$ versus $h$ we
also expect the LLQ parameters $K_q$ and $v_q$ to be sensitive to $h$. This
is shown in Fig.\ \ref{fig:gamma-h}b) and c) for the same wave
vector and temperatures as in panel a). At $T=0$ the LLQ parameters
can be obtained from Bethe Ansatz (BA)
\cite{Bogoliubov1986}. Regarding our approach it is very
satisfying to see that $K_q$ and $v_q$ almost quantitatively agree with
BA, and moreover that the remaining discrepancies diminish as
$T\rightarrow 0$ in the QMC approach. We emphasize that it should come as
no surprise, that expectation values obtained from {\it static} QMC agree
with other exact methods. Rather the main point is that $K_q$ and $v_q$
within our QMC approach follow from fits to a {\it dynamic} structure
factor. We note that the downturn of $v_q$ with $h$ is consistent with the
physical picture of a cosine band being filled completely as $h\rightarrow
h_c$.

Figure \ref{fig:gamma-q-h} displays $\gamma_q$ versus momentum for two
temperatures $T=0.1$, $0.25$ $J$ and various magnetic fields $h=0$, $0.5$,
$1.0$, up to half of the saturation field. This figure shows, that
increasing $h$ decreases the slope of $\gamma_q$ and modifies the curvature
as well. However, $\gamma_q(h>0)$ is not less then $\gamma_q(h=0)$ for large
momenta. To gauge if the introduction of $\gamma_q$ improves the agreement
between the renormalized LLQ and the imaginary-time QMC data, as compared
to the free LLQ, also for finite magnetic fields, we display the difference
between $\chi(q,\tau)$ from Eq.\ (\ref{mats}) and that obtained from QMC 
in Fig.\ \ref{fig:diff-h-delta}a) and b). This figure clearly supports
$\gamma_q \neq 0$. Similar however to the zero-field case, the QMC data
also suggests that at low temperatures, additional, albeit smaller,
renormalizations beyond diffusion are present.

\begin{figure}[tb]
\includegraphics[width=0.95\columnwidth]{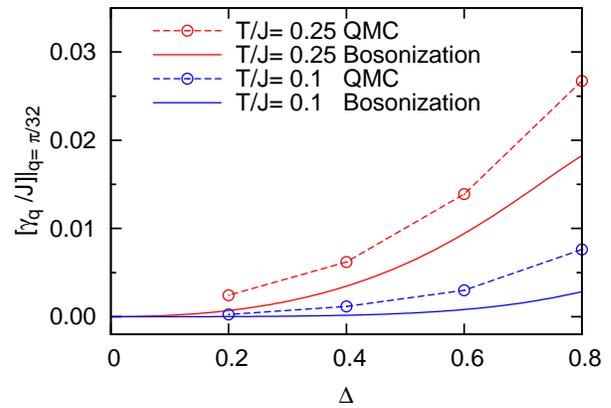}	
\caption[Anisotropy dependence of $\gamma_q$]
{(color online) Anisotropy dependence of the spin-current relaxation rate obtained from
bosonization (solid curves) and our method (symbols) for two temperatures $T/J= 0.1$, $ 0.25$.
For QMC, the first nonzero momentum $q=\pi/32$ on a system of length $L=64$ is considered.
}
\label{fig:gamma-del} 
\end{figure}

\section{Anisotropy}\label{Aispy}

Finally, we consider the role of finite anisotropy for the range of
$0<\Delta<1$. To this end Fig.\ \ref{fig:gamma-del} displays the relaxation
rate versus $\Delta$ for two temperatures $T/J=0.1$ and $0.25$ obtained from
two methods, i.e.\ our QMC approach and the perturbation theory based on
bosonization \cite{Sirker2009a,Sirker2011a}. For QMC the results are
plotted for the smallest, nonzero wave vector on a chain of length $L=64$,
i.e.\ $q=\pi/32$. The bosonization result is of the form $\gamma(T)=f(K,v)
T^{4K-3}$, where $f$ is a function of the Luttinger parameter $K$ and the
spin velocity $v$ which are exactly known from BA at zero temperature. This
figure shows, that the QMC results behave qualitatively very similar to the
findings from bosonization yet, QMC is consistently larger than the
latter. The figure does not show the point $\Delta=1$, since at that point
a similar comparison has already been performed in
Ref.\ \onlinecite{Grossjohann2010a}. The evolution of $\gamma_q$ with $\Delta$ as
the latter approaches zero is very much consistent with the fact that at
$\Delta=0$ the HAFC is in the XY limit, i.e.\ a free Fermi gas, which
displays fully ballistic spin transport, i.e.\ $\gamma=0$.

As for the case of finite magnetic fields, in Fig.\ \ref{fig:K,v-del} we now
turn to the dependence of the two LLQ parameters $K_q$ and $v_q$ for the
smallest wave vector on a $L=64$ site chain. Once again, values for these
parameters at $T=0$ are known from BA, i.e.\
\begin{equation}\label{eq:k,v-del}
K=\frac{\pi}{\pi-\arccos\Delta},\;v=\frac{\pi\sqrt{1-\Delta^2}}{2 \arccos\Delta} \,\, .
\end{equation}
In Fig.\ \ref{fig:K,v-del} results from the fits to QMC are compared with
BA. As for the finite field we observe very good agreement, which improves
as $T\rightarrow 0$.

\begin{figure}[tb]
\includegraphics[width=0.95\columnwidth]{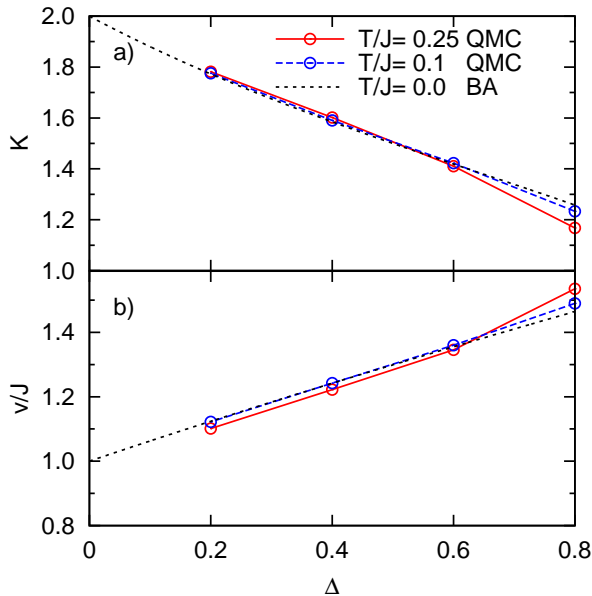}
\caption[Anisotropy dependence of $K$ and $v$]
{(color online) a) Luttinger parameter and b) spin velocity at finite temperature obtained from
QMC in terms of anisotropy for the first nonzero momentum at chain length
$L=64$. Luttinger parameter and spin velocity at zero
temperature known from BA are also shown for comparison purposes.}
\label{fig:K,v-del} 
\end{figure}

\begin{figure}[tb]
\includegraphics[width=0.95\columnwidth]{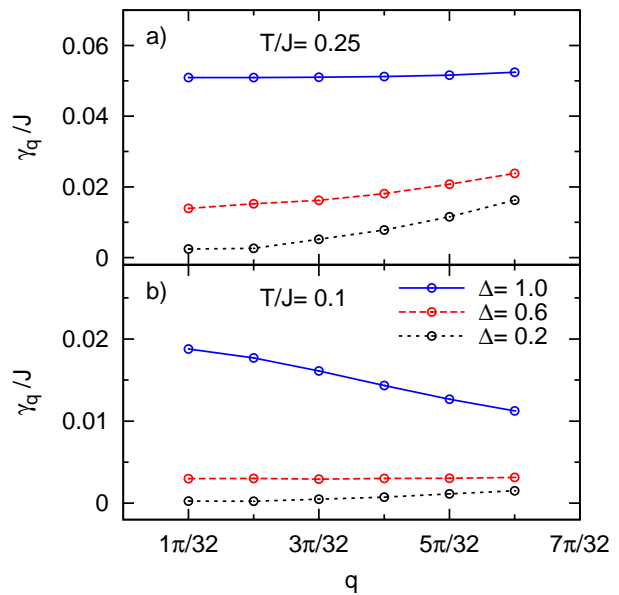}
\caption[Momentum dependence of $\gamma_q$ (anisotropic chain)]
{(color online) Momentum dependence of the spin-current relaxation rate $\gamma_q$ for
different anisotropies $\Delta=1.0, 0.6, 0.2$ at two temperatures a) $T=0.25 J$ and b) $T=0.1 J$.
The system-size here is set to $L=64$.}
\label{fig:gamma-momentum-del} 
\end{figure}

As for the previous sections, we also consider the effects of anisotropy on
the momentum dependence of the relaxation rate. This is shown in
Fig.\ \ref{fig:gamma-momentum-del} at two temperatures $T=0.1$, $0.25$ $J$
for the isotropic point $\Delta=1$ and two values of anisotropy
$\Delta=0.2$, $0.6$. Similar to the finite-field case, the slope and
curvature of $\gamma_q(\Delta)$ versus $q$ varies with $\Delta$, albeit
less strong. For the higher temperature, i.e.\ $T=0.25$ $J$, only a weak
momentum dependence can be observed at the isotropic point and an
increasing function in the anisotropic case. For $T=0.1$ $J$, $\gamma_q$
decreases with $q$ at $\Delta=1$, while for the anisotropic case, this
behavior is reversed.

The quality of the QMC fits to Eq.\ (\ref{eq:chiimag}) for a finite
$\Delta=0.8$ are depicted in Fig.\ \ref{fig:diff-h-delta}. They exhibit the
same trend as in all similar comparisons discussed in this work, i.e.\ a
finite $\gamma$ clearly improves agreement with QMC as compared to
$\gamma=0$, with however systematic deviations still present for very low
temperatures and finite momenta.

\section{Summary}\label{chainsum}

In summary, we have explored the dynamic structure factor of the
antiferromagnetic Heisenberg chain at long wave lengths and finite
temperatures. For this we have used the response function of a Luttinger
liquid including self-energy corrections which were parametrized in terms of
only a few relevant parameters. We have then determined these parameters by
fitting to high-precision imaginary-time quantum Monte-Carlo
calculations. Quite generically we find, that the self-energy renormalizes
the low-energy, long wavelength density spectrum of the Luttinger liquid
from a propagating `sound' mode into a diffusion mode. Apart from the
standard Luttinger parameters, the diffusion mode introduces one additional
parameter, i.e.\ a current relaxation rate.  We have analyzed this
relaxation rate with respect to several physical variables: temperature,
momentum, magnetic field, and anisotropy. Moreover, we have shown
consistency of our approach by performing finite-size analysis, checking
for anomalous diffusion, and comparing with exact diagonalization and
results from Bethe Ansatz. Our findings are qualitatively consistent with
previous studies using perturbation theory and bosonization, as well as
with a limited QMC analysis, considering only the role of temperature and
momentum for a single system-size. Future work should enlarge the
variational space for the self-energy, in order to capture corrections to
purely diffusive behavior.\\
{\it Note added:} After completion of this work we have become aware of
very recent results \cite{Karrasch2014z}, obtained in a different context,
with different methods, which are also consistent with a finite
zero-frequency limit of the regular spin conductivity in the HAFC.

\begin{acknowledgments} 
We thank P.\ Prelov\v{s}ek, X.\ Zotos, and J.\ Herbrych 
for very helpful discussions. 
Part of this work has been supported by the Deutsche Forschungsgemeinschaft
through FOR912, grant no.\ BR 1084/6-2, the European Commission through
MC-ITN LOTHERM, grant no.\ PITN-GA-2009-238475, and the NTH School for
Contacts in Nanosystems. W.B.\ thanks the Platform for Superconductivity and
Magnetism Dresden, where part of this work has been performed, for its kind
hospitality.
\end{acknowledgments}


\begin{thebibliography}{99}
\bibitem{fixbc}
E.g. at $\Delta=1$, $b=g^{2}/4-g^{3}(3-8\pi^{2}/3)/32+\sqrt{3}T^{2}/ \pi$ and
$c=g^{2}/4-3g^{3}/32-\sqrt{3}T^{2}/\pi$ in powers of the running coupling
constant $1/g+\ln(g)/2= \ln(\sqrt{\pi/2}\exp(G+1/4)/T)$ and
$G\approx0.577216\ldots$ is Euler's constant \cite{Lukyanov98}.
\end{thebibliography}
\end{document}